# Superhard and Superconducting $B_6C$

Kang Xia [a], Mengdong Ma [b], Cong Liu [a], Hao Gao [a], Qun Chen [a], Julong He [b], Jian Sun [a,*], Hui-Tian Wang [a], Yongjun Tian [b], and Dingyu Xing [a]

[a] *National Laboratory of Solid State Microstructures, School of Physics and Collaborative Innovation Center of Advanced Microstructure, Nanjing University, Nanjing 210093, China*

[b] *State Key Laboratory of Metastable Materials Science and Technology, Yanshan University, Qinhuangdao 066004, China*

**ABSTRACT**

Crystal structure searching and *ab initio* calculations have been used here to explore low-energy structures of boron carbides under high pressure. Under pressures of 85–110 GPa, a metastable $B_6C$ with $R\bar{3}m$ symmetry is found to be energetically more stable than the mixture of previous $B_4C$ and elemental boron. This $B_6C$ is a rhombohedral structure and contains mooncake-like $B_{24}$ clusters with stuffing of $C_2$ pairs. The mechanical and dynamical stabilities of this structure at ambient pressure are confirmed by its elastic constants and phonon dispersions. The bulk modulus and shear modulus of this newly predicted $B_6C$ at ambient pressure reach high values of 291 GPa and 272 GPa, respectively. The Vickers hardness is calculated to be around 48 GPa, and its melting temperature at ambient pressure is estimated to be around 2500 K, indicating that this metastable $B_6C$ is a potential superhard material with very good thermal stability. Interestingly, this superhard $B_6C$ structure is also found to be superconducting and its superconducting critical temperature ($T_c$) is evaluated to be around 12.5 K at ambient pressure by electron-phonon coupling.

\* Corresponding author.

*E-mail address*: jiansun@nju.edu.cn

# 1. Introduction

Boron and a large number of boron-rich materials exhibit outstanding mechanical, electronic and refractory properties, which makes them widely used in electronic and engineering fields [1–6]. Boron phases have complex bonding configurations such as two- and three- centers bonds, which lead to many allotropes composed of $B_{12}$ icosahedron, small interstitial clusters, and even fused supericosahedra [4,7,8]. The complexity of boron also brings about the richness of low-dimensional polymorphic structures, including two-dimensional (2D) sheets and one-dimensional (1D) chains for both elemental boron and boron-rich compounds [9–12]. These structural and bonding variations result in diverse properties such as superhardness and superconductivity [6].

To overcome the shortcoming of diamond which reacts with ferrous materials in cutting [3], searching for new superhard materials (Vickers hardness surpasses 40 GPa) [13] with good chemical inertness is in demand. Boron and boron-rich carbides possess outstanding mechanical properties and chemical inertness, making them to be widely used as cutting and coating materials [5,14]. The hardening mechanism of materials has been studied for over half a century [13,15-17]. The mechanical bulk modulus and shear modulus were found to be related to the hardness of materials [17,18]. In $RuB_2$, $OsB_2$, and $ReB_2$ synthesized with the high pressure and high temperature (HPHT) method [17], a linear dependence of Vickers hardness on shear modulus $G$ was observed, where $G$ is found to be sensitive to bond strength and directionality. $B_{12}$-icosahedral $B_4C$ is measured to have high value of bulk modulus and shear modulus of 247 and 200 GPa, respectively [18]. And its Vickers hardness is measured to be 32–35 GPa [19]. Theoretically, the hardness of covalent crystals is found to be closely related to the bond length, electronic density, degree of covalent bonding, bond metallicity and orbital hybridization [7,15,16]. For instance, taking into consideration the ionicities of boron-boron bonds in $B_{12}$ icosahedra of $α$-boron, the hardness of typical boron-rich solid $B_{13}C_2$, is validly calculated to be 44 GPa [7]. It agrees well with the experimental value of 45 GPa for $B_{13}C_2$ [20].

The hard $B_4C$ and superhard $B_{13}C_2$ can be synthesized under ambient pressure. For example, $B_4C$ can be obtained with hot-pressing at temperatures above 2200 °C [21] and $B_{13}C_2$ observed at the sintering temperatures of 1900 $^{\circ}$C [22]. Besides, some superhard carbon-rich borides were recently synthesized using a laser-heating diamond anvil cell under the HPHT condition. For instance, the synthesized cubic $BC_5$ was found to have a high bulk modulus of 335 GPa and a high Vickers hardness of 71 GPa [23]. The synthesized diamond-like d-$BC_3$ should have a bulk modulus of 350 GPa and a Vickers hardness of 53 GPa [24,25].

Elemental boron exhibits rich structural phase transitions under the HPHT condition [4,26–28], and its ground state structure at ambient condition was predicted to be $\tau$-B phase [28]. It should be noticed that some metastable phases of boron possess interesting electronic properties, such as superconductivity of $\alpha$-Ga structure [27] and superhard semiconductor of $\gamma$-$B_{28}$ phase [4]. According to the conventional Bardeen-Cooper-Schrieffer (BCS) theory [29], light atoms can usually promote the superconducting critical temperature ($T_c$). The superconductivity of elemental boron has been studied for 15 years. Eremets et al. [30] found that the nonmetal $\beta$-B transforms to a superconducting phase under about 175 GPa with $T_c$ = 6 K. The transition was later suggested to be related to the breaking of supericosahedra packing unit in a pressure-induced amorphization of $\beta$-B phase by the XRD experiment [31]. Using density functional theory calculations, the superconducting phase was suggested to be the $\alpha$-Ga structure [32,33]. Because the boron-doped diamond was discovered to be the type-II superconductor, boron-doped carbons and carbon-rich borides attracted more attention [34,35]. Increasing boron concentrations may raise their superconducting critical temperature, which can even exceed that of $MgB_2$ [2]. For example, the diamond-like cubic $BC_5$ (c-$BC_5$) was predicted to be a superconductor with a critical temperature of 45 K [36], which was synthesized later at 24 GPa and 2200 K with a laser-heated process in a large-volume multianvil apparatus [23]. An orthorhombic iron boride $FeB_4$ was discovered to be a superconductor with superhardness [37,38], which was later questioned by magnetization, resistance and load-dependent hardness measurements [39]. Interestingly, a newly identified beryllium hexaboride structure $\alpha$-$BeB_6$ was predicted to be both superhard and superconducting [40]. However, most of the known boron-rich carbides are insulators or semiconductors, which restrict their applications in some aspects. Whether there is a metallic or even superconducting superhard boron-rich carbide is still an open question.

## 2. Methods

In this work, we have extensively explored the B-C system by applying *ab initio* random structure searching (AIRSS) [41,42], a new boron-rich carbide ($B_6C$) is predicted to possess both superhardness and superconductivity, which serves as a candidate for multifunctional materials. This $B_6C$ phase is energetically more favorable than the $B_4C$ and c-$BC_5$ under high pressure. And its mechanical and dynamical stability at ambient pressure are checked by elastic constants and phonon spectra, which make this $B_6C$ phase to be a metastable phase. Here we searched for thermodynamically stable boron-rich carbides under high pressures using the AIRSS-technique [41,42] and the CASTEP code [43]. AIRSS has been a powerful tool to predict new structures theoretically including CO, and $N_2$ [44,45]. The energy stability of all the best candidate structures were verified by the projector-augmented wave (PAW) method [46,47] as implemented in the VASP code [48]. We employed the hard-version PAW Perdew-Burke-Ernzerh (PBE) [49] generalized gradient approximation (GGA) potentials for boron and carbon with a kinetic energy cutoff of 1050 eV, the Brillouin zone is sampled with a k-spacing of 2π×0.03 Å$^{-1}$. Due to the reason that the van der Waals (vdW) corrections seem to be important to describe the Boron element [28], the Grimme's DFT-D3 vdW corrections with the Becke-Jonson (BJ) damping [50,51] were taken into consideration in this work. The mechanical moduli were computed by VASP using the Voigt averaging [13,52–54]. The Vickers hardness was further calculated by the model presented by Chen et al. [55], rechecked carefully by the method presented by Guo et al. [16]. The electronic structures were calculated with the WIEN2k code [56], applying the linearized augmented plane wave (LAPW) method [57]. The Fermi surfaces were generated by using the WIEN2k and rendered by the XCysDen [58]. The electronic localizations were displayed by the visualization for the electronic and structural analysis (VESTA) tool [59]. Phonon dispersions and phonon density of states were studied by the PHONOPY code [60], combined with the VASP. We employed the QUANTUM-ESPRESSO code [61] within a density-functional perturbation theory (DFPT) approach [62] to calculate the Eliashberg spectral function $α^2F(ω)$ and the electron-phonon coupling constant ($λ$) [63] using ultrasoft pseudopotentials with a kinetic energy cutoff of 1088 eV and a 9×9×9 *k*-mesh. The thermal stability and melting point are calculated by *ab initio* molecular dynamics (*AIMD*) simulations in the *NpT* [74, 75] and *NVE* ensembles [76], respectively.

## 3. Results and Discussion

*3.1 Energy stability*

Recently, the $\tau$-B phase [28] has been predicted to be more stable than the $\alpha$-$B_{12}$ by the quantum mechanics simulations with the lattice zero-point vibrational energy (ZPE) and Grimme's dispersion corrections [50]. Both the above study [28] and our own works [65,66] indicate that the vdW corrections are important for molecular, layered and clustered structures, such as structures with $B_{12}$ clusters for boron carbides. Therefore, the Grimme's DFT-D3 vdw correction method with the BJ damping [50,51] is employed in this work. The formation enthalpy of B–C compounds is defined as,

$$h(BxCy) = [H(B_xC_y) - xH(B) - yH(C)] / (x + y), \qquad (1)$$

where $H(B_xC_y)$, $H(B)$, and $H(C)$ are enthalpies for $B_xC_y$ formula unit, stable elemental boron ($\alpha$-$B_{12}$, $\gamma$-$B_{28}$, and $\alpha$-Ga phases), and graphite or diamond-structured carbon respectively. The enthalpy difference between various compounds is calculated by equation mixed with energetically stable element under certain pressures, i.e.

$$\Delta H(B_xC_y) = [H(B_{x/y}C) + mH(B) + nH(C) - H(B_4C)] / d, \qquad (2)$$

where $n = 0$, $m = 4 - x/y$, $d = 1 + x/y$, if $x/y > 1$; $m = 4 - x$, $n = 1 - y$, $d = 4 + y$, if $x/y < 1$. Here $H(B_{x/y}C)$ represents the enthalpy for one $B_xC_y$ formula unit if $x/y < 1$.

As shown in Fig. 1a), the formation enthalpy sequence of three elemental borons ($\alpha$-$B_{12}$, $\gamma$-$B_{28}$ and $\alpha$-Ga) is in good agreements with the theoretical results in Ref. [3]. At pressures <18 GPa, the $B_4C$ framework tends to be formed from pure boron and diamond carbon. The formation enthalpy confirms the metastability of the previously known $BC_3$ (layered hexagonal) and $BC_5$ (diamond-like) [64].

We find that the $B_6C$ compound predicted in this work, as shown in Fig. 1a), possesses the lower formation enthalpy than that of the metastable layered hexagonal $BC_3$ and diamond-like $BC_5$ [23] under the pressure in a wide range of 22–200 GPa, which have been calculated to transform into experimentally observed diamond-like $BC_3$ and $BC_5$ [23,25], respectively. Although the lowest formation enthalpy of $B_6C$ is calculated to be positive, around 82 meV/atom under pressure of 50 GPa. However, the enthalpy of $B_6C$ is about 7 meV/atom lower than that of experimentally synthesized $BC_3$ and 22 meV/atom lower than the well-known $B_4C$ at 100 GPa, as shown in Fig. 1b). This indicates that $B_6C$ may have chance to be synthesized in HPHT experiments. For instance, the mixture of $B_4C$ and elemental boron may form $B_6C$ under pressures of about 85–110 GPa in the reaction of $B_4C + 2B \rightarrow B_6C$. While add diamond, $B_6C$ may decompose into $BC_3$ and boron under high pressures >110 GPa with the chemical reaction of $B_6C + 2C \rightarrow BC_3 + 5B$. It is known that the HPHT conditions are necessary to overcome the energetic barriers for the synthesis of B–C compounds, such as $B_4C$, $B_{13}C_2$, d-$BC_3$, and c-$BC_5$ [21–24].

Since the formation energy of $B_6C$ is the lowest one under 50 GPa, we discuss the temperature effects on the Gibbs free energy ($E_G$) at this pressure point. $E_G$ of electronic and phonon vibrations at a constant volume with a temperature range of 0–2000 K is calculated by the PHONOPY [60] using the quasi-harmonic approximation (QHA) [67]. The formation Gibbs free energy ($\Delta E_G$) at 50 GPa is calculated by the similar Eq. (1). The black solid line in Fig. 1c) shows that above 500 K, the temperature has larger influences on the $\Delta E_G$ of $B_6C$ structure than that below 500 K. $\Delta E_G$ decreases from 71 meV/atom at T = 0 K to about 42 meV/atom at T = 2000 K. The 11 meV energetically lower at T = 0 K than the formation enthalpy energy in Fig. 1a), is contributed by the ZPE difference between $B_6C$ and elements. $E_G$ of $B_6C$ calculated by formula (2) keeps almost 90 meV/atom lower than that of $B_4C$ under 100 GPa in a wide temperature range of 0–2000 K, indicating the energetical advantages over synthesized $B_4C$ phase and the possible reaction of $B_4C + 2B \rightarrow B_6C$ under the HPHT condition. The simulated XRD results of $B_6C$ (blue curve) and $B_4C$ (red one) are exhibited in Fig. 1d). The strongest diffraction peak of $B_6C$ at around 8.5° shifts about 0.3° from that of $B_4C$.

*3.2 Crystal structure, ELF and Fermi surface results*

The space group of the predicted $B_6C$ phase is $R\bar{3}m$ (No. 166). Its hexagonal unit cell can be formed by three rhombohedral primitive cells, as shown in Figs. 2a) and b).

This unit cell contains six $B_6C$ formula units, with $a = b = 7.668$ Å, $c = 5.337$ Å, $\alpha = \beta = 90°$, and $\gamma = 120°$. Boron atoms occupy two different Wyckoff positions in unit cell, i.e. B1 (18h): −0.1138 −0.2275 0.7693; B2 (18f): −0.3333 0.0349 0.3333; and carbon atoms occupy C (6c): 0.0000 0.0000 0.3482. Those boron atoms nearest to carbon atoms occupy B1 site. The 3-fold axis of primitive cell of $B_6C$ stay along two carbon atoms, viewed from Fig. 2b). Mooncake-like $B_{24}$-cluster is stuffed with such carbon pair in this structure, where every four boron atoms (half of them is B1) form a configuration of rhombus (purple solid lines drawn in Figs. 2b) and c)). Optimized by using PBE–D3 at 0 GPa, the bond length of C–C, B1–B2 and C–B1 are around 1.621, 1.784, and 1.636 Å, respectively.

The formation of these bonds are analyzed by three-dimensional electron localization function (ELF) and its contour plots in the (001) and ($1\bar{1}0$) planes, depicted in Fig. 3a and b). Viewed from the isosurface with a value of 0.75 $e$/Bohr$^3$ in Fig. 3a), the valence electrons of B and C atoms localize in directions along C–C, B1–B2 and C–B1 bonds. The values of ELF range from 0 to 1, where 0 represents a low electron density area and 1 indicate the perfect localization of electrons. Clearly, the ELF in sliced planes of (001) and ($1\bar{1}0$) exhibited in Fig. 3b), illustrate the localization of valence electrons in real place. The highest ELF value reaches about 0.94 in the areas between B1 and C, reflecting the strong covalent interaction of B1–C and C–C bonds. The localization along B1–B2 and B1–B1 bonds is weaker, with an ELF value of about 0.8. The weakest localization appears between two B2 atoms, with an ELF value of about 0.75. Nonlocalized electrons and metallic states occupy the blue and green areas, distributed around the $B_{24}$ clusters and $C_2$ pairs. Therefore, the $R\bar{3}m$ (166) boron subcarbide has strong covalent bonding interaction, although the overall bulk property belongs to metallic.

To discuss the mechanical stability of the $B_6C$ structure, we study whether symmetry-related elastic constants fulfill mechanical stable conditions or not. For trigonal symmetry (restricted to classes 32, $\bar{3}m$, 3m [68]), one has six independent elastic constants, i.e. $C_{11} = C_{22}$, $C_{33}$, $C_{12}$, $C_{13} = C_{23}$, $C_{14} = C_{56} = -C_{24}$, $C_{44} = C_{55}$; $C_{66} = (C_{11} - C_{12})/2$, all other $C_{ij} = 0$. The elastic constants of $B_6C$ calculated shown in Table 1 satisfy the mechanical stability conditions under hydrostatic pressure $P$ [54] ($C_{44} - P > 0$, $C_{66} - P > 0$, $C_{11} - C_{12} - 2P > 0$, and $(C_{13} - P)(C_{11} + C_{12}) - 2(C_{13} + P)^2 > 0$). The positive phonon dispersions calculated under ambient pressure and 100 GPa, as shown in up and down part of left panel in Fig. 5 respectively, demonstrate the dynamical stability of this rhombohedral $B_6C$ phase. We further study the thermal stability of $B_6C$ by performing *ab initio* molecular dynamics (*AIMD*) calculations, as shown in Fig. 6 a,b). The statistically averaged $B_6C$ structure maintains intact in the entire *AIMD* simulations for 12 ps using the *NpT* ensemble [74, 75] at ambient pressure and high temperature of 2400 K, as shown in Fig. 6 a). The closest C–C, B–B, and C–B bond lengths stay at around 1.60, 1.80, and 1.65 Å respectively, indicating the bond stability in the $B_{24}$ cluster and $C_2$ pairs. Thus the $B_6C$ structure can be obtained at ambient pressure and high temperatures. Using the *Z* method [76], we estimate its melting temperature to be abound 2500 K at ambient pressure, as shown in Fig. 6 b).

After discussions on the stability of $B_6C$ structure, we now focus on its electronic band structures and Fermi surfaces, as shown in Figs. 4 and 3c), respectively. The two bands crossing the Fermi energy level in the energy interval of about ±2 eV (colored as seen from Fig. 4), are dominantly contributed by the B1 (named in Fig. 2a) and b)) $p_z$, B2 $p_y$ and C $p_x + p_y$ states. The band dispersions near Fermi level are pretty strong in the whole Brillouin zone, except a small part on the *L*–Γ direction. This dispersionless band near the *L* point at the Fermi level should lead to a Van Hove singularity, which gives large cylinder-like Fermi surfaces around the *L*–Γ, as shown in the Fig. 3 c). This flat-dispersive band along the *L*–Γ line mainly comes from the B1 $p_z$ and C $p_x + p_y$ hybrid states. The $B_6C$ structure under pressure of 100 GPa owns very similar electronic bands, shown in the lower panels in Fig. 4, which demonstrates pressure has pretty small effect on the electronic structures of this $B_6C$ phase and also indicates that this phase is very stiff.

*3.3 Vickers hardness calculations*

It is noticed that elemental boron, carbon and their compounds are widely used as superhard materials for their excellent mechanical properties [18–20,23–25,69–71]. As one of boron-rich carbides, $B_6C$ is expected to possess excellent mechanical property, such as high *B* and *G*. Based on the Voigh-Reuss-Hill averaging [13], the corresponding *B* and *G* are derived from the calculated elastic constants [52–54]. If the *c/a* ratio does not change, the bulk modulus and shear modulus of hexagonal and trigonal symmetry [13,52–54] are $B = \frac{1}{9}(2C_{11} + 2C_{12} + 4C_{13} + C_{33} + 3P)$, bulk modulus under isotropic pressure *P* [52]; and $G = \frac{1}{5}(G^v_{eff} + 2C_{44} + 2C_{66})$, shear modulus at ambient pressure [53], respectively. $G^v_{eff}$ represents the energy per unit volume in a grain along its axis of symmetry. A uniaxial shear strain of unit magnitude is applied to the grain. While for hexagonal and trigonal symmetry, $G^v_{eff} = \frac{1}{3}(C_{11} + C_{33} - 2C_{13} - C_{66})$, $G^v_{eff} = \frac{1}{2}(C_{11} - C_{12})$, respectively. For the computations of Vickers hardness, here we applied two different models from Chen et al. [55] and Guo et al. [16].

The resulted elastic constants, mechanical moduli and corresponding Vickers hardness of predicted hexagonal $B_6C$ under 0 GPa, compared with the experimental values *Hv*, are exhibited in Table 1, compared with the quantities of *α*-$B_{12}$, diamond and other boron carbides. The structure details after PBE-D3 optimizations at 0 GPa, are listed in Table 2. The elastic constants and moduli calculated by PBE-D3 (BJ) agree with those of experimental values for the compounds studied. Compared with the experimental Vickers hardness *Hv*, the estimated value from Chen et al.'s model (*Hvc*) is slightly lower than that from Guo et al.'s method. This might due to the reason that boron carbides have very complicated bonding and electronic properties, such as strong covalent bonds and metallicity. The calculated *B* and *G* of $B_6C$ at 0 GPa are 291 GPa and 272 GPa, respectively. The Vickers hardness is calculated to be around 48.6 GPa from Guo *et al.*'s model, and 46.1 GPa from Chen *et al.*'s model. Both results suggest the $B_6C$ predicted in this work is a potential superhard material.

## 3.4 *Electron-phonon coupling effect*

The van Hove singularity and cylinder-like Fermi surface along the *L*–Γ direction inspired us to explore the electron-phonon coupling and the possible BCS superconductivity of $B_6C$. As shown in Fig. 5, a large range of the spectrum are dominated by the vibrations of B atoms. The contribution made by C atoms mainly comes from the C-projected DOS at $\omega$ of 700–940 cm$^{-1}$. The $\alpha^2F(\omega)$ curve (solid line) and the frequency-dependent electron-coupling ($\lambda(\omega)$) (dashed line) are exhibited in the right part of up panels in Fig. 5. Combined with $\alpha^2F(\omega)$, $\lambda(\omega)$ shows that the electron-phonon coupling is distributed among all of the phonon branches, excepting for low frequency modes at $\omega < 167$ cm$^{-1}$. Superconductivity has also been found in other boron-rich compounds, such as MgB$_2$ [2,72] and FeB$_4$ [37,38]. And the superconductivity in these systems are closely related to the electron-phonon coupling of their special Fermi surfaces and phonons. Here we find that the electron-phonon coupling constant of B$_6$C is almost distributed on all the phonon branches homogeneously. $\lambda$ is obtained to be around 0.69 by Riemann Sum of the frequency-dependent electron-phonon coupling function, in agreement with the results 0.70 by numerical integration of $\lambda(\omega)$ up to $\omega = \infty$, the $\omega_{\log}$ is calculated to be $\approx$ 364 K. Here $\omega_{\log}$ is defined as, $\omega_{\log} \equiv \lim_{n \to 0} \overline{\omega}_n = \exp(\frac{2}{\lambda}\int_0^\infty \frac{d\omega}{\omega}\alpha^2F(\omega)\ln\omega)$, where $\overline{\omega}_n$ represents the nth sequence of average frequencies $\overline{\omega}_n \equiv \langle \omega^n \rangle^{1/n}$. According to the Allen-Dynes modified McMillian expression for $T_c$ [73]: $Tc \equiv \frac{\omega_{\log}}{1.2}\exp[-\frac{1.04(1+\lambda)}{\lambda - \mu^*(1+0.62\lambda)}]$, together with a commonly used effective Coulomb repulsion parameter $\mu* = 0.10$, we estimated the $T_c$ of B$_6$C at 0 GPa to be around 12.5 K. The electron-phonon properties at 100 GPa are also computed, as shown in down panels of Fig. 5. Although the $\lambda$ is computed to be 0.76, which is increased a little compared with that at 0 GPa. However, the $\omega_{\log}$ is decreased to 246.0 K. And the $T_c$ of B$_6$C at 100 GPa is estimated to be around 10.4 K. Using the adjustable $\mu* = 0.13$, $Tc$ of B$_6$C is calculated to be around 9.6 and 8.3 K at pressure of 0 and 100 GPa, respectively.

## 4. Conclusion

Using crystal structure searching based on *ab initio* calculations [42], a metallic boron subcarbide B$_6$C [space group: $R\overline{3}m$ (No. 166)] is found. The thermodynamic stability, mechanical properties and the electron-phonon coupling of this $B_6C$ phase are studied throughout. This $B_6C$ structure consists of mooncake-like $B_{24}$ clusters stuffing with dumbbell-like $C_2$ pairs. is predicted to be energetically more stable than mixture of boron carbide ($B_4C$) and elemental boron under HPHT, and can be metastable at ambient pressure. Its elastic constants are found to fulfill the mechanical stability requirements of this rhombohedral structure. And the phonon spectra of this structure, both at 0 and 100 GPa, show its dynamical stability. This $R\bar{3}m$ $B_6C$ is also thermally stable with a high melting point of about 2500 K. The calculated Vickers hardness of this $B_6C$ at 0 GPa reaches a high value of about 48 GPa and its superconducting critical temperature is estimated to be around 12.5 K and 10.4 K at 0 and 100 GPa, respectively. These results indicate this newly predicted $B_6C$ is a superhard and superconducting material, which might have potential applications under some particular circumstances.

## Acknowledgments

J.S. is grateful for the financial support from the National Key Projects for Research & Development of China (Grant No. 2016YFA0300404), 973 project (Grant No. 2015CB921202), the National Natural Science Foundation of China (Grant Nos. 51372112 and 11574133), the NSF of Jiangsu Province (Grant No. BK20150012), the Fundamental Research Funds for the Central Universities (No. 020414380068/1-1), and Special Program for Applied Research on Super Computation of the NSFC-Guangdong Joint Fund (the second phase) under Grant No. U1501501. Part of the calculations was performed on the supercomputer in the High Performance Computing Center of Nanjing University and "Tianhe-2" in the NSCC-Guangzhou.

**Figures**

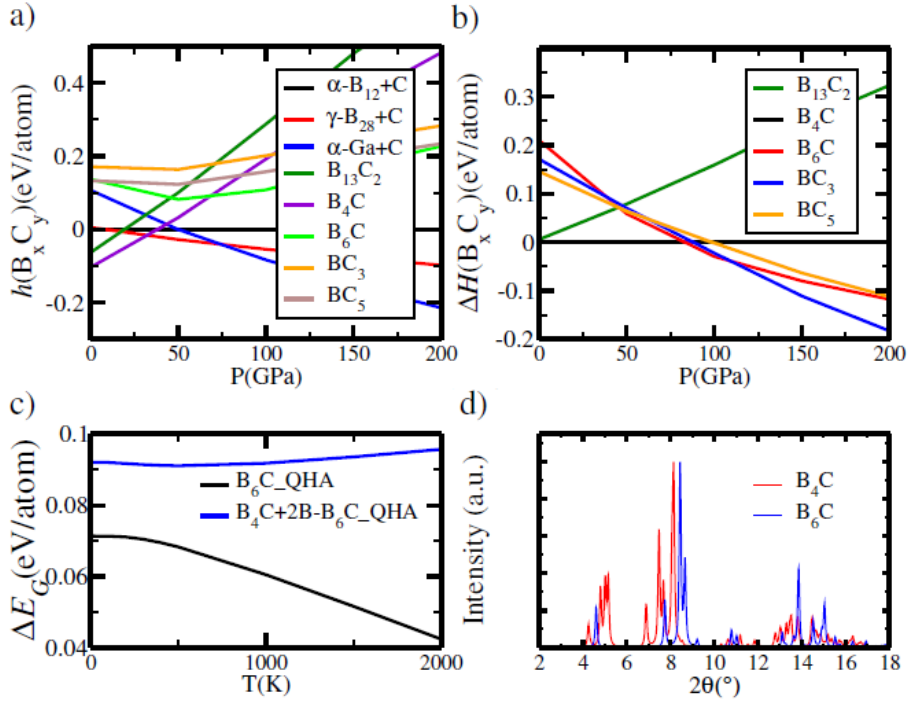

**Fig. 1.** Energy-pressure relations for all $B_xC_y$ structures under pressures in a range of 0–200 GPa. a) The formation enthalpy curves relative to α-$B_{12}$ plus carbon (data at 0 GPa is calculated from graphite while data at other pressure points are from diamond). b) Enthalpy-pressure relations relative to $B_4C$ plus elemental boron or carbon. c) The formation Gibbs free energy versus the temperature. Blue solid line in c) represents the formation Gibbs free energy from quasi-harmonic approximation (QHA) ($\Delta E_G$) of $B_4C$ plus α-Ga boron related to that of $B_6C$ under 100 GPa. Black solid line is Gibbs free energy $E_G$ of $B_6C$ under 50 GPa, related to that of γ-$B_{28}$ plus diamond. d) is the calculated XRD pattern of the newly predicted $B_6C$ (blue) with a wavelength of 1.54056 Å, compared to the known $B_4C$ (red). All structures were optimized using PBE–D3 (BJ) [49–51], including our predicted $B_6C$, other known elemental borons (α-$B_{12}$, γ-$B_{28}$, α-Ga) [4] and diamond phase, two boron-rich carbides ($B_{13}C_2$, $B_4C$) [18,20] and two carbon-rich borides ($BC_3$, $BC_5$) [64].

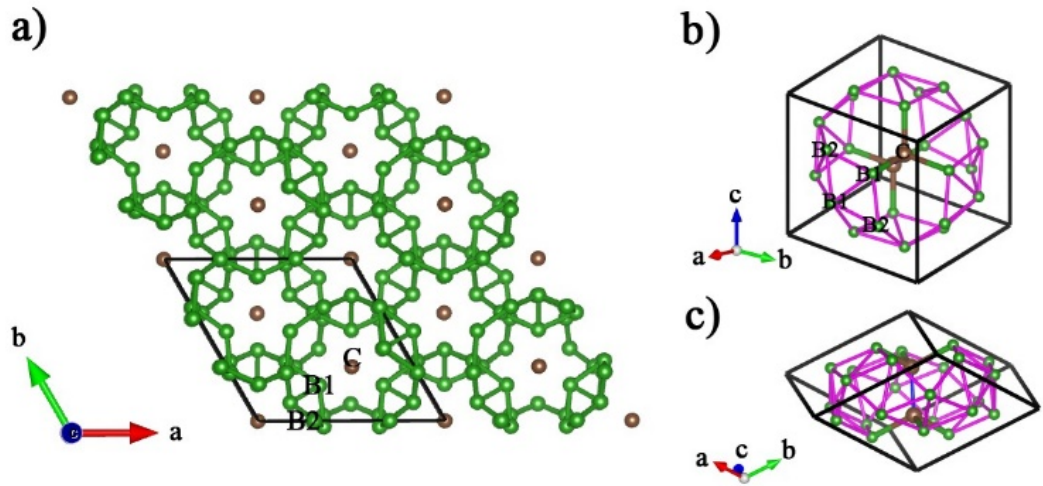

**Fig. 2.** Crystal structure of $B_6C$: (green for B atoms and gray for C atoms) for a) (2 × 2 × 1) supercell of unit cells, and b) and c) two different views of corresponding primitive cell. B1 and B2 label boron atoms occupying two different Wyckoff positions. C represents that of carbon atoms.

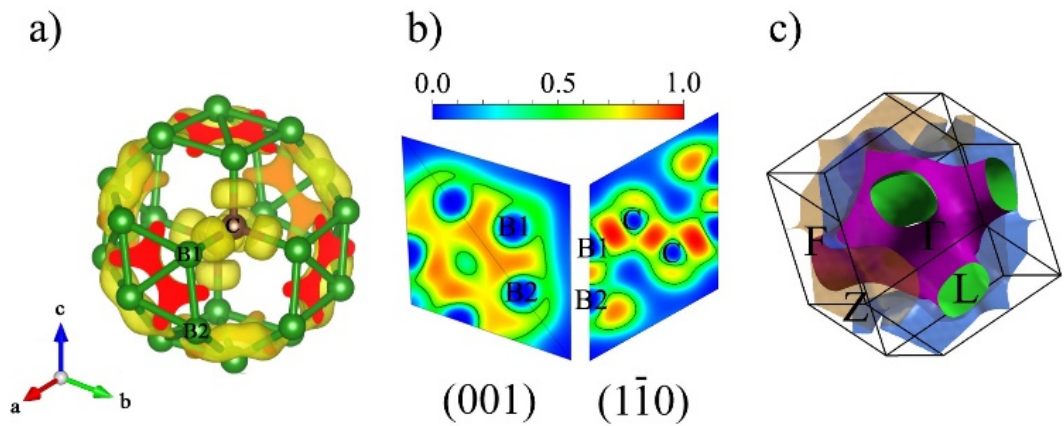

**Fig. 3.** The electron localization function (ELF) and Fermi surfaces of $B_6C$ under 0 GPa. a) Three-dimensional ELF with isosurface value of 0.75 $e$/Bohr$^3$ of $B_6C$. b) The ELF contour plot in the (001) and ($1\bar{1}0$) planes, where yellow and blue areas represent gaining and losing electrons, respectively. c) The calculated Fermi surfaces of $B_6C$. We splice (001) plane and adjacent slice through four B2 atoms to observe the ELF between two B2 atoms, as shown in left part of b).

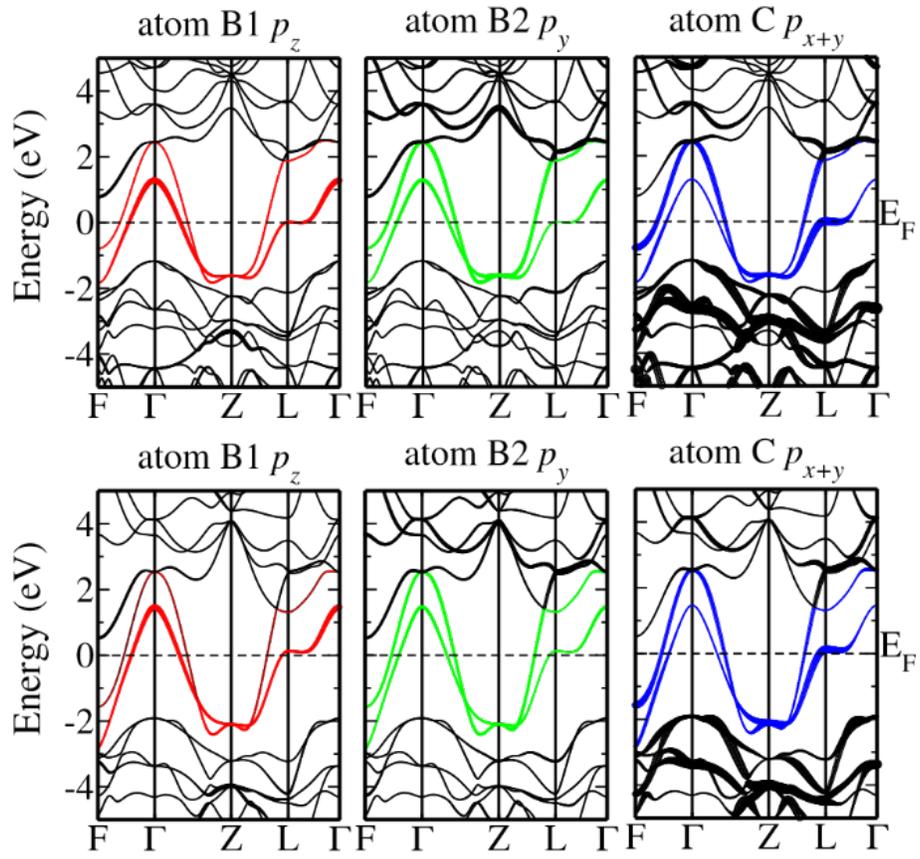

**Fig. 4.** The calculated band structures of $B_6C$ at 0 GPa (up panels) and 100 GPa (down panels). Colored lines characterize the main contributions from B1 $p_z$ (red in left part), B2 $p_y$ (green in the middle) and C $p_x+p_y$ (blue in the right) orbits to bands crossing the Fermi energy level ($E_F$) (zero energy dashed-line).

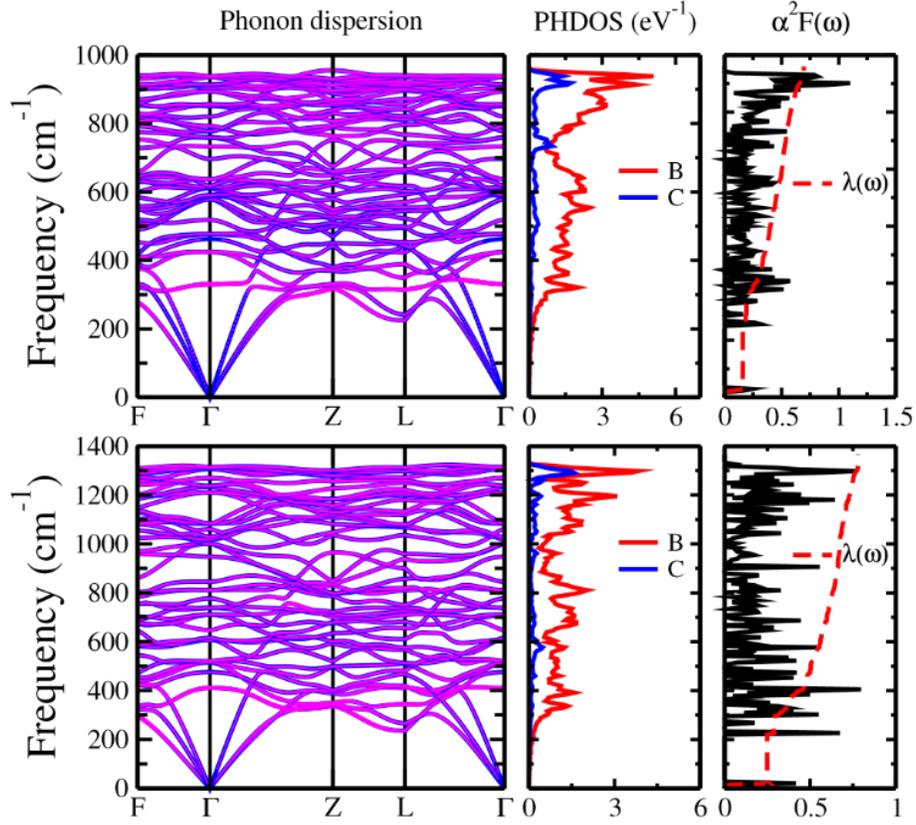

**Fig. 5.** Electron-phonon properties of $B_6C$ under 0 GPa (up panels) and 100 GPa (down panels). Left: Phonon dispersions. Middle: Atom projected phonon DOS. Right: The Eliashberg spectral function $\alpha^2F(\omega)$ (solid line) and frequency-dependent electron-phonon coupling constant $\lambda(\omega)$ (dashed line). The size of magenta circles around the phonon dispersion curves displays the phonon width.

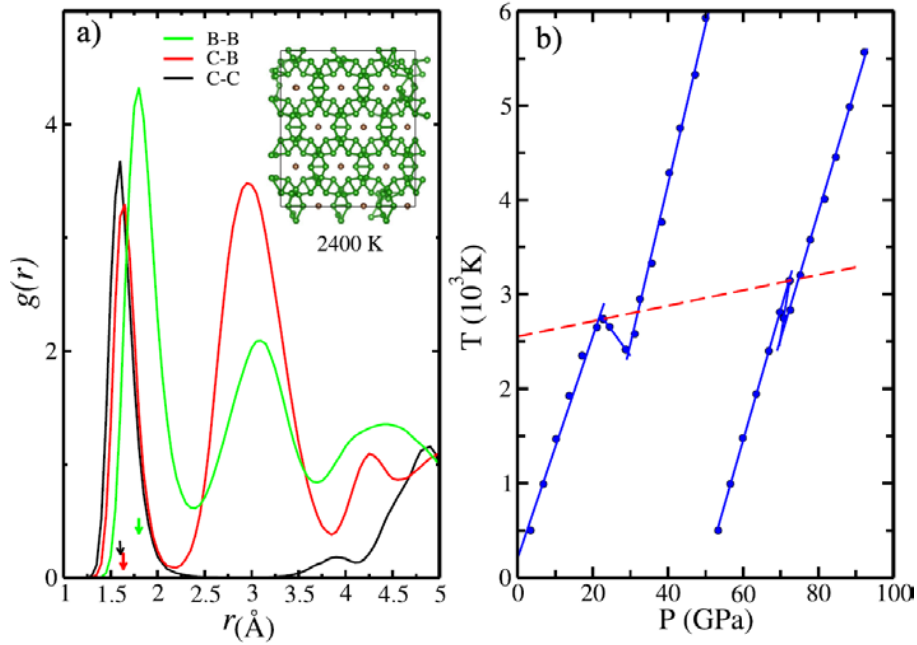

**Fig. 6.** a) Radial distribution functions (*g(r)*) for the C–C, B–B, and C–B separations during the entire *AIMD* simulations for $R\bar{3}m$ B$_6$C at ambient pressure and temperature of 2400 K. The black, green, and red arrows represent the average C–C, B–B, and C–B distances, respectively. Inset is the statistically average B$_6$C supercell. The *NpT* ensemble with a Langevin thermostat [74,75] is employed in the simulation running for 12 ps. b) The melting temperature at ambient pressure (around 2500 K) estimated using *Z* method for B$_6$C by *AIMD* simulations with the *NVE* ensemble [76]. All MD simulations are performed using a $\sqrt{2}\times2\sqrt{2}\times2$ supercell of B$_6$C with 336 atoms.

**Table 1.**

The elastic constants ($C_{ij}$ (GPa)), bulk and shear moduli ($B$ and $G$ (GPa)) and Vickers hardness of $\alpha$-$B_{12}$, diamond and boron carbides from our DFT calculations under ambient pressure, compared with others' works. The elastic constants in this work were calculated by the VASP code [48] employing PBE–D3 (BJ) methods [49–51]. $Hvc$ and $Hvg$ represent the Vickers hardness estimated by the model of Chen et al. [55] and Guo et al. [16], respectively. The calculated Vickers hardness is compared with the experimental values, which are denoted with $Hv$.

| Compounds | Work | $C_{11}$ | $C_{12}$ | $C_{13}$ | $C_{33}$ | $C_{44}$ | $C_{66}$ | $B$ | $G$ | $Hvc$[55] | $Hvg$[16] | $Hv$ |
|---|---|---|---|---|---|---|---|---|---|---|---|---|
| Diamond | this work | 1068.8 | 126.8 | 126.8 | 1068.8 | 565.9 | 565.9 | 441 | 546.9 | 99.9 | 93.4 | - |
| | experiment | 1079±5[a] | 124±5[a] | - | - | 578±2[a] | - | 443[b] | 535[b] | - | - | 96±5[b] |
| $\alpha$-$B_{12}$ | this work | 509.7 | 119.9 | 48.5 | 646.8 | 216.9 | 194.9 | 233.4 | 222.4 | 41.6 | 40.2 | - |
| | experiment | - | - | - | - | - | - | 224[c] | - | - | - | 42[d] |
| $B_4C$ | this work | 487.2 | 157.2 | 152.2 | 509.8 | 170.1 | 227.5 | 267.5 | 190.0 | 25.9 | 39.2 | - |
| | experiment | - | - | - | - | - | - | 247[e] | 200[e] | - | - | 32-35[f] |
| $B_{13}C_2$ | this work | 545.8 | 137.6 | 83.7 | 488.0 | 104.3 | 204.1 | 243.3 | 167.5 | 22.8 | 45.9 | - |
| | experiment | - | - | - | - | - | - | 231[e] | 189[e] | - | - | 45[d] |
| d-$BC_3$ | this work | 669.8 | 192.3 | 192.3 | 669.8 | 400.5 | 400.5 | 351.5 | 368.1 | 63.9 | 43.6 | - |
| | theory [25] | 658.4 | 194.7 | - | - | 392.5 | - | - | - | - | - | 52.5 |
| c-$BC_5$ | this work | 870.2 | 174.1 | 86.9 | 1027.3 | 386.6 | 348.1 | 384.8 | 385.6 | 62.3 | 65.4 | - |
| | experiment [23] | - | - | - | - | - | - | 335±8 | - | - | - | 71±8 |
| $B_6C$ | this work | 569.4 | 160.7 | 118.6 | 684.6 | 340.6 | 204.4 | 291.0 | 272.1 | 46.1 | 48.6 | - |

[a]Reference(Ref.) [69]. [b]Ref. [70]. [c]Ref. [71]. [d]Ref. [20]. [e]Ref. [18]. [f]Ref. [19].

**Table 2.**

The crystal structure details for newly predicted $B_6C$ and other boron carbides ($B_{13}C_2$, $B_4C$, d-$BC_3$, and c-$BC_5$), including the $\alpha$-$B_{12}$ and diamond phases. ITN is the International Crystallographic Table Number.

| Compound | Space group (ITN) | Lattice parameters (Å) | Wyckoff positions |
|---|---|---|---|
| Diamond | $Fd\bar{3}m$ (No. 227) | a = b = c = 3.559 | C(1) 8a: -0.2500 -0.2500 0.2500 |
| $\alpha$-$B_{12}$ | $R\bar{3}m$ (No. 166) | a = b = 4.862, c = 12.463 | B(1)18h: -0.4287 -0.2143 -0.2249 |
| | | | B(2)18h: -0.2717 -0.1359 -0.3577 |
| $B_4C$ | Cm (No. 8) | a = 8.720, b = 5.573, c = 5.023, $\beta$ = 61.25° | B(1) 4b: 0.4407 -0.2433 -0.8027 |
| | | | B(2) 4b: 0.1648 -0.1611 -0.9879 |
| | | | B(3) 4b: -0.4412 -0.2457 -0.1984 |
| | | | B(4) 4b: -0.1673 -0.1625 -0.0089 |
| | | | B(5) 2a: 0.1912 0.0000 -0.3154 |
| | | | B(6) 2a: -0.1918 0.0000 -0.6845 |
| | | | B(7) 2a: -0.0069 0.0000 -0.3414 |
| | | | B(8) 2a: -0.4982 0.0000 -0.4920 |
| | | | C(1) 2a: 0.0049 0.0000 -0.6786 |
| | | | C(2) 2a: 0.3801 0.0000 -0.6079 |
| | | | C(3) 2a: -0.3734 0.0000 -0.3846 |
| $B_{13}C_2$ | $R\bar{3}m$ (No. 166) | a = b = 5.625, c = 12.027 | B(1)18h: 0.1617 0.3235 -0.3582 |
| | | | B(2)18h: 0.1054 0.2108 -0.1135 |
| | | | B(3) 3b: 0.0000 0.0000 -0.5000 |
| | | | C(1) 6c: 0.0000 0.0000 -0.3803 |
| d-$BC_3$ | $I\bar{4}3m$ (No. 217) | a = b = c = 7.309 | B(1) 8c: 0.3974 -0.3974 0.3974 |
| | | | B(2) 8c: 0.2435 -0.2435 0.2435 |
| | | | C(1)12e: 0.2678 0.0000 0.0000 |
| | | | C(2)24g: 0.1180 -0.3757 -0.3757 |
| | | | C(3)12d: 0.5000 0.0000 -0.2500 |
| c-$BC_5$ | P3m1 (No. 156) | a = b = 2.545, c = 6.374 | B(1) 1a: 0.0000 0.0000 0.0068 |
| | | | C(1) 1a: 0.0000 0.0000 0.2599 |
| | | | C(2) 1b: 0.3333 0.6667 0.3384 |
| | | | C(3) 1b: 0.3333 0.6667 0.5818 |
| | | | C(4) 1c: 0.6667 0.3333 0.6643 |
| | | | C(5) 1c: 0.6667 0.3333 0.8997 |
| $B_6C$ | $R\bar{3}m$ (No. 166) | a = b = 7.668, c = 5.337 | B(1)18h: -0.1138 -0.2275 0.7693 |
| | | | B(2)18f: -0.3333 0.0349 0.3333 |
| | | | C(1) 6c: 0.0000 0.0000 0.3482 |